# Two Approaches for Programming Education in the Domain of Graphics: An Experiment

Luca Chiodini[a] , Juha Sorva[b] , Arto Hellas[b] , Otto Seppälä[b] , and Matthias Hauswirth[a]

a    Software Institute, Università della Svizzera italiana, Lugano, Switzerland
b    Department of Computer Science, Aalto University, Espoo, Finland

**Abstract**

Context    Graphics is a popular domain for teaching introductory programming in a motivating way, even in text-based programming languages. Over the last few decades, a large number of libraries using different approaches have been developed for this purpose.

Inquiry    Prior work in introductory programming that uses graphics as input and output has shown positive results in terms of engagement, but research is scarce on whether learners are able to use programming concepts learned through graphics for programming in other domains, transferring what they have learned.

Approach    We conducted a randomized, controlled experiment with 145 students as participants divided into two groups. Both groups programmed using graphics in Python, but used different approaches: one group used a compositional graphics library named PyTamaro; the other used the Turtle graphics library from Python's standard library. Student engagement was assessed with surveys, and programming knowledge with a post-test on general programming concepts and programming tasks in the domain of graphics.

Knowledge    We find few differences between the two groups on the post-test, despite the PyTamaro group having practiced on problems isomorphic to those in the post-test. The participants traced a compositional graphics program more accurately than a 'comparable' turtle graphics program. Both groups report high engagement and perceived learning; both perform well on simple program-writing tasks to create graphics.

Grounding    Our findings are based on a controlled experiment with a count of 145 participants, which exceeds the sample size indicated by power analysis to detect a medium effect size. The complete instrument and teaching materials used in the study are available as appendixes.

Importance    This study adds further evidence that graphics is an engaging domain for introductory programming; moreover, it shows that the compositional graphics approach adopted by PyTamaro yields engagement levels comparable to the venerable turtle approach. Compositional graphics code appears to be easier to trace than turtle graphics code. As for conceptual knowledge, our results indicate that practicing on programming tasks isomorphic to those of the test can still not be enough to achieve better transfer. This challenges programming educators and researchers to investigate further which graphics-based approaches work best and how to facilitate transfer.

## The Art, Science, and Engineering of Programming

## 1 Introduction

Computing teachers are always on the lookout for new ways to engage novice programmers. One popular strategy to promote engagement is to have students create programs that produce fun graphics rather than dull textual outputs. Many programming environments, software libraries, microworlds, and other tools have been built to support these pedagogies (see, e.g., [32]). A well-established example is turtle graphics [41], now available as a library for many programming languages and even included in the standard library of languages popular for teaching such as Python. Another prominent example is the media-computation pedagogy introduced by Guzdial [22], who designed a curriculum based on manipulating sounds and images in order to teach computing concepts.

A number of studies have demonstrated that graphics-based pedagogies have positive effects on engagement [2, 23, 48, 53]. Something that has received comparatively little attention is whether students generalize the concepts they learn in the domain of graphics into programming concepts and transfer them to programming in other domains. Papert [40] argued that turtle graphics is an excellent vehicle to teach mathematics, programming, and problem-solving in general. However, Pea et al. [46] conducted studies with children using turtle graphics in Logo and did not find the hoped transfer. Planning skills did not improve after a year of programming in Logo [45]. And even for programming skills, the understanding of concepts depended highly on the context. As an example: "a child who had written a procedure using REPEAT which repeatedly printed her name on the screen did not recognize the applicability of REPEAT in a program to draw a square." [46]

Different approaches to graphics for introductory programming have been proposed, including libraries that treat graphics as immutable values to be composed. To the best of our knowledge, no study has been conducted to empirically evaluate which approach best fosters transfer from programming using graphics to programming in general.

Our study seeks insight into both engagement and conceptual transfer. We compare two fundamentally different approaches to graphics: compositional graphics and the well-established turtle graphics. As an example of the former approach, we use PyTamaro, a graphics library designed for beginner programmers by some of the authors. So far, the library has been only evaluated analytically, drawing on arguments from theory, prior empirical findings in computing education, and anecdotal evidence of PyTamaro itself [10].

We conduct a randomized, controlled experiment involving 145 participants. We ask the following research questions:

**RQ1** *Is there a difference in conceptual transfer from a short programming tutorial with a compositional graphics or a turtle graphics library to programming outside the domain of graphics?*

**RQ2** *After a tutorial following either approach, are there differences in how students read or write programs?*

**RQ3** *Do the two approaches lead to different levels of student engagement or perceived learning?*

## 2 Related Work

This section starts by exploring the multifaceted relationship between graphics and programming education. We then review different approaches to program graphics, comment on the few studies that empirically evaluate them, and argue why one specific approach is worth investigating for transfer of specific programming concepts.

### 2.1 Graphics and Programming Education

There are multiple senses in which programming education can be related to graphics. This section clarifies the scope of this work.

A fundamental separation exists between *graphical* and *text-based* programming. Doing 'graphical programming' commonly means using a visual programming language to create programs *graphically*, instead of *textually*. Scratch [33] is an example of a visual programming language popular in educational contexts at the school level. Scratch programs are composed of 'blocks' connected visually. The success of block-based visual programming languages led to the creation of Blockly [44], a library that facilitates the creation of other block-based languages. Scratch also inspired the creation of Snap! [26], a block-based language in which users can define their own blocks. Our work instead concerns *text-based* programming languages, which university-level educational contexts use more frequently.

Even when written using text-based programming languages, programs can involve 'the domain of graphics' to different extents. For instance, a program can produce 'static' graphics (2D or 3D), 'animated' graphics, or 'interactive' graphics (known as *Graphical User Interfaces*).

While creating interactive graphical programs has also been explored as a possibility to learn programming (e.g., [17, 34]), our work focuses on approaches to create static 2D graphics, among which there is already considerable variety.

### 2.2 Graphics Libraries for Beginners

Chiodini et al. [10] identified three families of 2D graphics libraries designed for teaching programming in a textual programming language: (1) canvas-based graphics; (2) turtle graphics; and (3) compositional graphics. We briefly describe each below, focusing on the latter two, which are directly relevant to the present study.

#### 2.2.1 Canvas-Based Graphics

Libraries in this family draw graphics by placing elements on a canvas. Elements can be freely positioned at locations defined by Cartesian coordinates relative to a global origin. These programs are imperative in nature and heavily reliant on mutable state. Learners frequently use 'magic numbers' that are not only hard-coded but that depend on each other implicitly. Examples include Python's `cs1graphics` [21], `acm.graphics`'s derivative Portable Graphics Library [51] for Java, as well as Processing [49], which is popular in the field of digital art.

### 2.2.2 Turtle Graphics

An alternative way to draw graphics is to control a 'turtle' that carries a 'pen.' [41] The program on the right of Table 1 uses the `turtle` library that comes standard with the Python language to draw (the black outline of) a house. This is a common introductory example in turtle-based pedagogy [42].

■ **Table 1** Drawing a house (adapted from Abelson et al. [1], originally from Papert et al. [42]) with PyTamaro and `turtle`

| Graphic | PyTamaro | Turtle |
|---------|----------|--------|

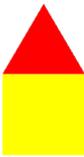

```
floor = rectangle(100, 100, yellow)
roof = triangle(100, 100, 60, red)
house = above(roof, floor)
show_graphic(house)
```

```
for _ in range(4):
    forward(100)
    right(90)
left(60)
for _ in range(3):
    forward(100)
    right(120)
```

In contrast to canvas-based libraries, there are no global coordinates. The perspective is *local*, relative to the turtle. Papert argued that turtle geometry is learnable because it is "body syntonic" [40]: the turtle's perspective is the same as the learner's. (We note however that most turtle libraries allow some form of positioning using global coordinates, partially going against the principle of body syntonicity.)

One of the goals of turtle graphics is to create abstractions that work as "building blocks" [1, 24]. From the program on the right in Table 1, it would be straightforward to grab a few lines as written and define, say, a procedure that draws a roof wherever the turtle happens to be. However, like canvas-based programs, this program also relies on mutable state: the command `forward(100)`, for example, depends on the turtle's current heading and position, and mutates the latter. Together, the turtle's position, heading, and pen status (up/down and color) constitute a global state that persists across procedures and that the programmer must attend to [10, 24].

The `left(60)` call in the Turtle code of Table 1 is a basic example of what Abelson et al. [1] termed "interface steps": additional code that modifies the global state in order to 'glue together' the solutions to subproblems. In this simple program, the subproblems are the single floor and the roof, and one needs an interface step to rotate the turtle in preparation for drawing the roof. Interface steps are a burden for the (beginner) programmer [24].

### 2.2.3 Compositional Graphics

A different approach for programming graphics was first proposed by Henderson [27] and later advanced by Finne et al. [20]. It is particularly popular among those who embrace functional programming; this is reflected in programming languages (such as Pyret [56]), libraries (such as Racket's image teachpack [3, 19]), and curricula (such as *How to Design Programs* [18]), which support and encourage this approach.

PyTamaro is a recent entry in this space, enabling beginners to do compositional graphics in Python. The program on the left of Table 1 shows how to draw the house with PyTamaro. The most striking difference to both canvas- and turtle-based programs is that this program treats graphics as values to be composed. Like the other libraries in this family, PyTamaro provides functions for creating basic shapes (e.g., `rectangle`) and for combining graphics (e.g., `above`). Another key difference is that the graphical values are immutable: one may compose them but not change them.

From this program, it would be easy to abstract a function that creates a roof: the solutions to the floor and roof subproblems are independent (apart from the width of the house, if one wishes to keep the floor and the roof at the same width). Those sub-solutions are explicitly combined (using `above`) to form a solution to the overall problem. There are no interface steps in the sense described above.

The PyTamaro program in Table 1 does not contain coordinates, and indeed the library features no global coordinate system. In fact, PyTamaro does not feature even *local* coordinates—e.g., coordinates relative to the top-left corner of a graphic—and differs in this from most libraries in the compositional graphics family. This is a facet of the library's minimalist design, which intentionally provides only a small number of features chosen to support conceptual learning.

### 2.3 Evaluations of Graphics-Based Approaches — And the Challenge of Transfer

Multiple studies report positive effects on *student engagement* from graphics-based pedagogies such as media computation [23, 48, 53] and turtle graphics [2]. After having accumulated a decade of experiences with media computation curricula, Guzdial [23] summarized their observations on their impact. The highlights of these findings include gains in student engagement, which led to a drastic reduction in failure rates (from 50 % to under 15 %). Moreover, female participation in the courses increased, stabilizing above 40 %.

A few studies have found evidence of transfer from graphics-based programming to *mathematics*. Noss [39] carried out an experiment showing that Logo helped to learn certain geometrical concepts. Schanzer et al. [52] presented initial data that show a measurable transfer of skills from a 'functional' programming curriculum (which also includes compositional graphics) to algebra when instructional materials are carefully aligned to the concepts normally covered in math classes.

Empirical evidence remains scarce on whether programming with graphics helps with *general programming skills*. Already in the 1980s, Pea et al. [46] looked into claims that Logo (and its turtle graphics) helps with problem-solving in general but found little evidence in support; moreover, they found that after 30 hours of programming with Logo, "children's grasp of fundamental programming concepts such as variables, tests, and recursion … was highly context-specific" [46], thus illustrating how difficult it is for learners to transfer their knowledge from a particular context or domain to others. More recently, Guzdial explored what they called the "learning hypothesis" of their media computation curricula, but the results were inconclusive or negative compared to a traditional curriculum [23].

### 2.4 Compositional Graphics Approaches Should Have Potential for Conceptual Transfer

Chiodini et al. [10] identified several trade-offs in graphics-library design for novices and argued that a compositional graphics library like PyTamaro can be a meaningful alternative to comparable libraries, under the right circumstances and given certain pedagogical goals.

As noted above, PyTamaro is designed to assist in the acquisition of fundamental programming concepts. For instance, few language constructs are needed for writing PyTamaro programs (e.g., no classes and objects), the API is minimal (e.g., no functions for loading external images), and neither mutable state nor coordinate-based operations are present. That is, PyTamaro has *deliberate limitations*: learners are not given access to certain functionality. This design should help manage complexity for beginner programmers, guide them towards better-quality programs, and nevertheless engage them meaningfully not only with graphics but with key computing content that is not specific to the graphics domain. In other words, PyTamaro is claimed to hold the potential for improved transfer.

The decomposition of problems into independent subproblems is key not only to professional programming [38, 43, 54] but broadly to computational thinking [58] and problem-solving [4]. However, learning decomposition and, relatedly, abstraction is hard [29, 37]. Several of PyTamaro's intended benefits involve these key concepts and skills. Mutable state and 'interface steps' (Section 2.2.2) in turtle graphics hinder effective problem decomposition. State makes it harder to reason about a subproblem in isolation: to understand what is the effect of a given piece of code, novices have to mentally reconstruct the state of the turtle. Lewis [31] documented how both school- and college-level students struggle with the turtle's state, leading to issues that are hard to debug. Compositional graphics approaches like PyTamaro eliminate this problem by offering only pure functions that produce immutable graphics.

Treating graphics as values to be composed also facilitates *visual decomposition* [10]. One may look at an image and identify its components (e.g., the roof and floor in our trivial example), and see how those visual components compose into an overall image; this maps directly to how the subproblems' programmatic solutions compose into an overall program. Learners may be guided to visually decompose graphics and thereby learn how to decompose programming problems and to compose the subprograms that solve them. Visual decomposition is relatively straightforward when objects are immutable and one does not need to consider components' locations as coordinates.

In the present work, we seek empirical evidence of the possible better transfer to programming in other domains when learning using a compositional graphics approach. We use the well-established and popular turtle graphics approach as a point of comparison.

## 3 Methodology

### 3.1 Procedure

We designed a randomized, between-subjects experiment with two conditions. In both conditions, the participants worked through a programming tutorial consisting of four 'mini-lessons' with a Python graphics library. We selected PyTamaro in its English API version as an example of a compositional graphics library (Section 2.2.3) and `turtle` from Python's standard library for turtle graphics (Section 2.2.2).

Since the study took place in a proctored computer laboratory and the participant count was high, we arranged four identical sessions over two weekdays. Each session consisted of four phases (Figure 1): a pre-survey, a teaching intervention phase, a post-survey, and a post-test. The teaching intervention was different for the two groups, as was part of the post-test (as explained below); the other phases were identical for both. We allowed participants a maximum of 90 minutes to complete the entire session.

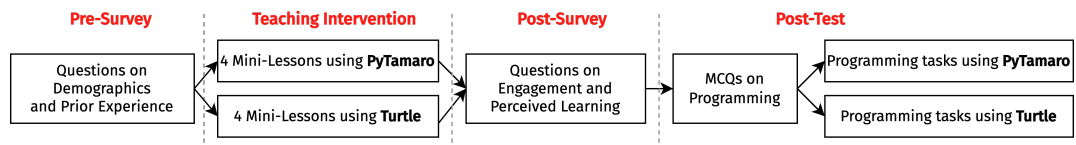

■ **Figure 1** The timeline of each session

Although some of the authors oversaw the sessions, they did not directly teach anything to either group. Instead, the intervention took the form of a self-paced tutorial on a sequence of web pages. This decision was made in an effort to increase the reproducibility of our findings and to eliminate biases in favor of our own library, PyTamaro.

**No Pre-Test?**   By design, our experimental setup did not include a pre-test. This means we cannot compute learning gains, but the decision can nevertheless be justified both methodologically and pragmatically.

First, the differences between the two groups are ironed out because we assigned the large number of participants to the two conditions randomly. This follows the recommendations of Campbell and Stanley: "While the pretest is a concept deeply embedded in [researchers'] thinking ... it is not actually essential to true experimental designs. ... the most adequate all-purpose assurance of lack of initial biases between groups is randomization" [5].

Second, a pre-test on programming could have brought about learning and muddled our results. There is evidence that just taking a test again leads to better learning outcomes [6]. We wanted to avoid that and instead study the effect of the teaching intervention with a compositional graphics library like PyTamaro, checking for transfer using the Turtle group as a baseline.

Third, the time constraints given by the context forced us to consider whether to have a pre-test or a longer teaching intervention. We decided to dedicate a greater fraction of the limited time of the experiment to the intervention.

### 3.2 Context and Participants

We recruited participants from an introductory programming course (CS1) at a large European research university that is not where PyTamaro originated and where PyTamaro had never been used before. The course is held during the first semester of undergraduate studies and targets non-majors in CS, especially students from other engineering fields. The course uses Python and adopts what might be described as a 'typical imperative programming pedagogy'; it does not feature any graphics-based programming.

None of the present authors are involved in teaching the course. The sessions took place during the third week of the course, outside class hours. During the first two weeks, the course had covered variables, basic I/O, assignment statements, simple arithmetic, and `if` and `while` statements; `for` loops were introduced in the third week. We organized the study very early in the course to limit prior programming knowledge. We ruled out the possibility of running the study even before the course start date: besides our participants having yet to begin their university path, it would have been unfeasible to cover meaningful content in a short teaching intervention with absolute beginners.

We advertised the study during the introductory lecture of the course. Participation was voluntary. Upon completion (but irrespective of performance), the participants were rewarded with a minor amount of course credit and a movie ticket.

A web platform provided all the materials and assessments and took care of randomization and data collection in accordance with the local anonymization policies. Each participant consented to the use of their anonymous data.

### 3.3 Pre-Survey

The participants answered a pre-survey with questions on three areas. First, we asked standard demographic questions. Second, we gauged their prior knowledge asking how many lines of code they had written before and whether they had ever programmed graphics. Third, we surveyed their attitude towards programming with three Likert items. The full questions are available in Appendix A.

### 3.4 Teaching Intervention

The teaching intervention for each group consisted of four 'mini-lessons.' Each such lesson took the form of a web page and consisted of text, executable snippets of Python code, and a few illustrations. Some of the snippets were ready to execute as-is, but the majority offered only a starting point for the participants to write their own code as instructed.

### 3.4.1 Interplay Between Pedagogy and Library

The lessons were designed purposely for this experiment. We tried to align the materials for the two conditions closely, while still adopting a meaningful use of each library. This was difficult, as there is an interplay between the tools one chooses and the pedagogy one may then adopt; each library is associated with a 'typical' pedagogy that tries to match its strengths and minimize its weaknesses.

At one extreme, a library might make a concept inaccessible because there is no support for it. For example, when graphics are not treated as values and functions are essentially procedures, as in turtle graphics, it is impossible to construct nested calls or other composite expressions with graphical values. At the other extreme, a library might make it practically mandatory to teach a certain concept. PyTamaro has, in common use, a parameterless function (`empty_graphic`) and various non-commutative functions that take two parameters (e.g., `above`), effectively forcing learners to deal with these concepts.

Somewhere in the middle of the spectrum, a library may nudge pedagogy towards certain concepts that are particularly compatible with it; for example, PyTamaro does not mandate exploiting associativity and multiple ways to decompose a particular problem, but it invites teachers and learners to explore these topics. Conversely, a library and its standard pedagogies may not particularly need a concept, but the concept may still be introduced despite not being prominent.

We sought a balance within these constraints to keep the experiment fair, especially avoiding favoring PyTamaro.

### 3.4.2 Lessons' Content

The lessons focused on using variables and functions, and on composition in general. Broadly, they emphasized expressions, an essential concept even in non-predominantly functional languages such as Python, but that is often neglected by traditional imperative pegagogies [9] like the one adopted in our CS1 course.

PyTamaro's approach is compositional and exploits expressions. It thus offers the right opportunities to explain these concepts, which were only briefly introduced in the first part of the CS1 course that took place before the experiment.

For both groups, we created materials in two natural language versions: one in English, another in Finnish. Each participant was free to choose whichever of the two; 39 % used the English version. The identifiers in Python code were identical (in English) in both language versions.

The first lesson introduced the idea of a software library and showed how to call library functions. For PyTamaro, functions' parameters and return values were visualized with a 'plumbing diagram' similar to Harvey [25, Ch. 2]. For Turtle, animated GIFs showed how the turtle executes a sequence of commands including movements and rotations.

The second lesson guided both groups toward drawing something slightly more interesting: the house from Table 1. The CS1 course had not yet covered function definitions, whose introduction from scratch would have required too much time. We opted to provide both groups with functions such as `square` and `triangle` and focus on their usage.

The third lesson explained how to draw a more complex graphic with two houses and a wall in between, to whet the learners' appetite for constructs that repeat computations. The PyTamaro group practiced combining variables and nesting, whereas the Turtle group focused on the importance of the order in the sequence of commands.

The fourth and final lesson introduced repeated computation with `for` loops, which students had just started practicing in the CS1 course. Both groups drew a 'street' of five houses side by side: the PyTamaro group used the parameterless `empty_graphic` function to initialize an 'accumulator' variable (cf. zero in summation), the Turtle group added an 'interface step' to reposition the turtle at the end of each iteration.

Here we only briefly outlined the contents of the lessons. The complete version for both groups is available in Appendix B.

### 3.5 Post-Survey

Immediately after the teaching intervention, we asked the participants to complete another survey. This was to explore RQ3 by eliciting their opinions on the lessons they just experienced, their level of engagement with programming in the domain of graphics, and whether they had actually perceived to be learning programming.

We formulated four hypotheses for engagement and three for perceived learning. The independent variables are (separately) the approach followed and the gender.

On engagement. *There is a difference in…*

**H3a** *… how interesting they think the tutorial was.*

**H3b** *… how fun they find programming graphics.*

**H3c** *… how much more they like programming graphics over programming in other domains.*

**H3d** *… how much they would like to learn more with graphics.*

On perceived learning. *There is a difference in…*

**H3e** *… how much they feel they have learned about programming.*

**H3f** *… how much they feel they already knew that approach to program graphics.*

**H3g** *… how much they feel they already knew the programming concepts taught.*

Each hypothesis corresponds to a seven-point Likert item in the post-test, answerable from "not at all true" to "completely true" (Table 8). The exact items as presented to the participants are shown in full in Appendix C.

### 3.6 Post-Test

In total, our post-test had nine questions whose themes are listed in Table 2. The post-test can be logically divided into two parts (even though this division was not visible to the participants).

The first part was identical for both groups. It consisted of six multiple-choice questions that relate to general programming concepts and RQ1, whose associate hypothesis is the following:

■ **Table 2** The themes of our post-test questions (Q1 to Q9) and how they map to the hypotheses derived from our first two research questions

| H | Post-Test | Topic / Task |
|---|---|---|
| H1 | Q1 | Use a variable more than once in an expression. |
| | Q2 | Nest function calls. |
| | Q3 | Use a function with more than one parameter. |
| | Q4 | Call a parameterless function. |
| | Q5 | Exploit an associative operation for multiple solutions. |
| | Q6 | Determine the initial value of a loop's accumulator variable. |
| H2a | Q7 | Tracing: Given a program, determine the size of the result. |
| H2b | Q8 | Writing: Create a program to draw a simple graphic. |
| H2c | Q9 | Modifying: Modify a given program that draws a graphic. |

**H1** *There is a difference in conceptual transfer between the group that followed a programming tutorial with PyTamaro or with Turtle, as measured on programming tasks outside the domain of graphics.*

We operationalize transfer as participants correctly answering the six multiple-choice questions.

The second part of the post-test relates to our second research question, from which we formulate three specific hypotheses:

*After following a programming tutorial with PyTamaro or with Turtle, there is a difference in how learners...*

**H2a** *... trace an existing program that creates a graphic.*

**H2b** *... write a program from scratch to create a graphic.*

**H2c** *... modify a given program that creates a graphic to adapt to new requirements.*

### 3.6.1 Q1 to Q6: Multiple-Choice Questions on Programming

The first six questions, Q1 to Q6, were multiple-choice questions unrelated to graphics. The questions targeted expression-related programming concepts (function calls, variable use, composition with operators or nesting) where the PyTamaro approach could yield better transfer, given that these concepts were only explicitly practiced in PyTamaro's teaching intervention. The Turtle group thus served as a baseline with respect to these questions: the Turtle participants had to answer these questions on the basis of whatever they had learned (or failed to learn) prior to the experiment.

Table 3 shows the alignment between the six multiple-choice questions and the examples featured in the teaching materials for the PyTamaro group. In the learning sciences literature, these are known as isomorphic tasks (or simply as isomorphs). We will comment further on transfer and its challenges in the Discussion (Section 5.3).

The table only summarizes the stems. For completeness and reproducibility, the questions with answer options appear in full in Appendix D.

■ **Table 3** Each multiple-choice question targeted an abstract concept that should be learned in the teaching intervention phase for the PyTamaro group and then transferred to an isomorphic task in the testing phase (Figure 2). For compactness, the tasks from the intervention (left) and the questions (right) are summarized here as single sentences. The full questions and answers can be found in Appendix D.)

<div align="center">

**Abstract Concept**

</div>

| Teaching Intervention with PyTamaro | Post-Test (Both Groups) |
|---|---|
| *Q1: The same variable can be used more than once in an expression.* | |
| `beside(house, house)` is valid | is `print(word + word + word)` valid? |
| *Q2: Function calls can be nested.* | |
| `rotate(45,` <br> `  rectangle(200, 100, green)` <br> `) is valid` | is `sqrt(` <br> `sqrt(16)` <br> `) valid?` |
| *Q3: A function can have multiple parameters and their order matters.* | |
| `above(ground_floor, roof)` <br> is valid and different from <br> `above(roof, ground_floor)` | is `subtract(10, 7)` <br> valid and different from <br> `subtract(7, 10)`? |
| *Q4: A function can have zero parameters and calling it still requires parentheses.* | |
| `empty_graphic()` is valid | is `fake_random()` valid? |
| *Q5: Multiple valid decompositions: if $\otimes$ is associative, $a \otimes (b \otimes c) \equiv (a \otimes b) \otimes c$.* | |
| `beside(house,` <br> `  beside(wall, house))` <br> is equivalent to <br> `beside(beside(house, wall),` <br> `  house)` | is `combine("re",` <br> `combine("stau", "rant"))` <br> equivalent to <br> `combine(combine("re", "stau"),` <br> `"rant")` ? |
| *Q6: The initial value of a loop's accumulator variable is the operation's neutral element.* | |
| when combining graphics, <br> initialize `result` to `empty_graphic()` | when multiplying numbers, <br> should `result` be initialized to 1? |

### 3.6.2 Q7 to Q9: Programming Tasks in the Graphics Domain

Unlike the questions on general concepts described above, the other three post-test questions could not be identical for the two groups, as the questions involve programming graphics and the two groups learned different approaches for that. We strove for tasks that are as close to each other as possible and yet respect the idiosyncrasies of each approach. Nevertheless, as the two groups' programs are not identical, this part of our results speaks not only of what the participants learned during the intervention but also of the characteristics of 'typical' code written using the two approaches to program graphics.

Below, we consider each of the hypotheses related to RQ2 in turn.

### 3.6.3 Tracing (H2a)

Question 7 asked participants to trace a program that draws four squares in a two-by-two grid, as shown in Table 4. The participants were prompted for the dimensions of the resulting drawing. An answer is considered correct only when both dimensions are correct.

This question checks for differences in how well participants can trace a program (hypothesis H2a).

■ **Table 4**  Q7: What are the width and height of the resulting drawing?

| PyTamaro | Turtle |
|---|---|
| ```
a = square(10, black)
b = square(10, black)
c = above(a, b)
d = square(10, black)
e = square(10, black)
f = above(d, e)
g = beside(c, f)
show_graphic(g)
``` | ```
pencolor("black")
square(10)
left(90)
forward(10)
right(90)
square(10)
forward(10)
square(10)
right(90)
forward(10)
left(90)
square(10)
``` |

### 3.6.4 Writing (H2b)

Question 8 was designed to address the second hypothesis related to RQ2: are there differences between the two approaches when learners write a program from scratch? The participants were asked to write a Python program to draw the simple graphic in Table 5. The PyTamaro group were expected to create a hammer's head and handle with the `rectangle` function, to compose them together, and to rotate the composite graphic. The Turtle group were expected to use a combination of movements and rotations to draw a colored letter T.

■ **Table 5**  Q8: Write a program to draw the given graphic. (One correct answer is shown for each group.)

| PyTamaro | | Turtle | |
|---|---|---|---|
| 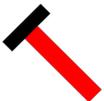 | ```
head = rectangle(120, 30, black)
handle = rectangle(40, 200, red)
hammer = above(head, handle)
rotated_hammer = rotate(45,
    ↪ hammer)
show_graphic(rotated_hammer)
``` | 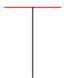 | ```
pencolor("red")
forward(200)
backward(100)
right(90)
pencolor("black")
forward(250)
``` |

### 3.6.5 Modifying (H2c)

Question 9 gave the participants a program that places five identical houses (like the one in Table 1) next to each other with a `for` loop. This program was the same as the one featured in the fourth mini-lesson.

■ **Table 6**  Q9: The participants were asked to double the dimensions of the houses.

| PyTamaro | Turtle |
|----------|--------|
| ```<br>ground_floor = square(100, yellow)<br>roof = triangle(100, 100, 60, red)<br>house = above(roof, ground_floor)<br>n_houses = 5<br>street = empty_graphic()<br>for i in range(n_houses):<br>    street = beside(street, house)<br>show_graphic(street)<br>``` | ```<br>n_houses = 5<br>for i in range(n_houses):<br>    pencolor("yellow")<br>    square(100)<br>    pencolor("red")<br>    left(60)<br>    triangle(100)<br>    right(60)<br>    forward(100)<br>``` |

Successfully modifying the PyTamaro program requires identifying the code that deals with the create-a-house subproblem, and updating the relevant arguments given to `square` and `triangle`. In contrast, modifying the Turtle program also requires the participant to position the turtle correctly before each loop iteration (e.g., editing the call to `forward`). A difference in the success ratio on this task between the groups would validate hypothesis H2c.

This task is meant to challenge the participants on a small scale with issues of maintainability. The turtle approach inherently has one additional challenge because the solutions to the subproblems, such as drawing a single house, cannot be composed independently of their specifics (the width of one house).

### 3.7 Analysis

Our study is mostly quantitative. We ran a power test to compute an appropriate minimal sample size, seeking to keep false positives under 5 % ($\alpha = 0.05$) and false negatives under 20 % ($\beta = 0.2$). We speculated on a "medium" effect size ($d = 0.5$) in either direction (two-tailed test). These constraints yielded a minimal sample size of 64 participants per group (cf. Table 2.4.1 in [12]); our participant count (145) exceeds this minimum.

We use parametric tests, trusting the Central Limit Theorem to guarantee normality on our large sample. We do not make assumptions about the direction of effects, or on the equality of variances. We therefore compare means with two-tailed, independent-samples $t$-tests without the equal-variance assumption—a.k.a. Welch's $t$-tests.

Following widespread recommendations [55, 57], we report each $p$-value together with an effect size (Cohen's $d$). Instead of performing corrections for multiple comparisons and then making claims of statistical significance, we consider $p$-values and effect sizes together in an attempt to interpret our results' real-world significance [55].

Below, we will note where a *p*-value is under the traditional threshold for significance $\alpha = 0.05$ or where an effect size is at least "small" ($d \geq 0.20$ as suggested by Cohen [11]).

Some of the questions in our pre- and post-surveys were on a seven-point Likert scale. We treat the responses to these questions as interval data, while acknowledging that there is no universal consensus on whether it is acceptable to do so [30]. (We adopted a seven-point scale, as simulations show that using more points brings the distribution closer to normal [59].) We thus use the parametric *t*-test for Likert items; the non-parametric Mann–Whitney–Wilcoxon test would also be appropriate, but it has been shown that the two tests have similar power for almost all distributions [15].

To complement our inferential statistics, we conducted semi-formal analyses of certain student responses (e.g., types of programming errors made), as described below under Results.

## 4 Results

As noted above, data was collected from four sessions spread over two weekdays. The sessions had 38, 39, 36, and 32 participants, respectively, for a total of 145 participants. The web platform we used automatically assigned participants to groups with equal probability. We ended up with 70 participants in the PyTamaro group and 75 in the Turtle group.

The PyTamaro group spent an average of 47 minutes on the whole session, which is somewhat longer than the Turtle group's 42 ($d = 0.38$, $p = 0.02$). We hypothesized that this might be due to some participants' prior exposure to programming with turtle graphics (Section 4.1), so we separately checked only those participants who had never programmed any graphics before; this reduced the difference to roughly three minutes (47 vs. 44; $d = 0.26$, $p = 0.19$).

### 4.1 Pre-Survey

The participants had an average age of 22 years (standard deviation 5.0). 71 participants (49 %) identified as female, 70 (48 %) as male, 0 as non-binary, 2 as other, and the remaining 2 preferred not to disclose gender information.

The vast majority of participants (120, 83 %) reported to have completed the first three rounds of exercises in the CS1 course they were taking. Pre-CS1 experience with programming was uncommon, with some exceptions.

65 participants (45 %) reported having written 0 lines of code before CS1 (excluding any HTML and CSS). 33 (23 %) reported fewer than 50 lines in total, 33 (23 %) fewer than 500, 11 (8 %) fewer than 5 000, and 3 (2 %) over 5 000. When asked whether they had ever written a program that draws graphics, 104 (72 %) participants answered no, 29 (20 %) yes, and the remaining 12 (8 %) were not sure. (We did not directly ask the students which tools they used for programming with graphics, but given the setting, we expect that Scratch is at the top of the list, and that some would also have done turtle graphics; the compositional graphics approach is likely to be very rare.)

Below, we use the term *novice* for participants who both had written fewer than 500 lines of code and had never programmed graphics before. By this measure, 107 participants (74 %) were novices; they ended up evenly split across the two groups, with 53 novices assigned to PyTamaro and 54 to Turtle.

As expected, given the large sample and randomization, there was no significant difference in self-reported prior programming knowledge between the groups ("lines of code written" on a scale between 0 and 4; $p = 0.59$, $d = -0.08$).

The participants generally expressed positive attitudes to programming. On a scale from 1 to 7, they strongly agreed that "it is useful for me to know programming," with an average rating of 6.21($\pm$0.84). Moreover, they moderately agreed that "programming is fun" (5.34 $\pm$ 1.36) and did not agree that "programming is boring" (2.25 $\pm$ 1.27).

### 4.2 Transfer to Programming Concepts (H1)

We rated the answers to the six multiple-choice questions, Q1 to Q6, as either incorrect (including "I don't know") or correct. Table 7 shows the results as percentages of correct answers for each question.

■ **Table 7**  Proportions of correct answers on Q1 to Q6, shown first for all participants and then for novices only

| | | Q1 | Q2 | Q3 | Q4 | Q5 | Q6 |
|---|---|---|---|---|---|---|---|
| **All** | PyTamaro % ($N = 70$) | 82.9 % | 80.0 % | 97.1 % | 57.1 % | 94.3 % | 68.6 % |
| | Turtle % ($N = 75$) | 81.3 % | 78.7 % | 93.3 % | 58.7 % | 90.7 % | 73.3 % |
| | Delta % | 1.5 % | 1.3 % | 3.8 % | -1.5 % | 3.6 % | -4.8 % |
| | *p*-value | 0.81 | 0.84 | 0.29 | 0.85 | 0.41 | 0.53 |
| | Effect size | 0.04 | 0.03 | 0.18 | -0.03 | 0.14 | -0.10 |
| **Novices** | PyTamaro % ($N = 53$) | 79.2 % | 81.1 % | 96.2 % | 56.6 % | 92.5 % | 60.4 % |
| | Turtle % ($N = 54$) | 81.5 % | 74.1 % | 94.4 % | 51.9 % | 90.7 % | 70.4 % |
| | Delta % | -2.2 % | 7.1 % | 1.8 % | 4.8 % | 1.7 % | -10.0 % |
| | *p*-value | 0.77 | 0.39 | 0.67 | 0.63 | 0.75 | 0.28 |
| | Effect size | -0.06 | 0.17 | 0.08 | 0.09 | 0.06 | -0.21 |

Overall, the participants fared decently well on these questions, which targeted basic expression-related programming concepts: average correctness across questions and groups was 80 %. Q3 (multi-parameter functions) and Q5 (equivalence of associative solutions) were solved correctly by more than 90 % of participants. Q4 (calling a parameterless function) and Q6 (initialize a variable for a loop) proved to be the hardest, with aggregate averages of 58 % and 71 %, respectively.

The differences are negligible on all six questions. Only three questions meet the very low bar of 0.10 for effect size: Q3 and Q5 in favor of PyTamaro and Q6 in favor of Turtle. None of these differences are statistically significant.

As shown in the bottom half of the table, we also looked separately into novice performance. Again, we found no major differences. The only effect size to pass the "small effect" threshold was −0.21 in favor of Turtle on Q6, and none of these results are statistically significant either.

We checked for a correlation between the time participants spent on the post-test and their performance on the multiple-choice questions. However, this correlation was effectively zero: $r^2 = 0.02$.

### 4.3 Programming Tasks (H2)

The last three questions in the post-test targeted the three facets of RQ2. We present the results for each related hypothesis in turn in the next subsections.

#### 4.3.1 Tracing (H2a)

In the PyTamaro group, 90 % of the participants answered their tracing task correctly, whereas only 37 % of the participants in the Turtle group correctly answered their 'comparable' task. The difference is statistically significant ($p < 0.001$), and the effect size is very large ($d = 1.29$).

We also looked at novices separately and found a similar trend. 89 % of PyTamaro novices answered correctly, compared to only 33 % for Turtle. Again, the difference was statistically significant ($p < 0.001$) and the effect size large ($d = 1.36$).

#### 4.3.2 Writing (H2b)

Both groups performed rather well on this task. 59 out of 70 (84 %) in the PyTamaro group solved the task correctly (i.e., the program does what was asked), as did 63 of 75 (84 %) in the Turtle group. Looking to understand the student programs in more detail, we semi-formally analyzed the incorrect answers.

The eleven incorrect answers in the PyTamaro group can be characterized as follows. Three participants did not submit a solution; two tried to rotate the individual graphics before composing them (which leads to issues related to the bounding box); one supplied `rotate`'s arguments in the wrong order; one rotated the hammer in the wrong direction; one had a typographical error; and three had a mix of other issues.

The turtle group had twelve incorrect answers: four participants mixed rotation (e.g., `left`) with movement (e.g., `forward`); two drew the letter 'upside down'; two used wrong lengths; one colored the entire drawing in red; one used strings where integers were called for (e.g., `"250"`); one drew an extra line; and one issue we were unable to classify.

#### 4.3.3 Modifying (H2c)

On this question, too, both groups performed rather well. 56 out of 70 (80 %) in the PyTamaro group solved the task correctly, and in the Turtle group the number was even higher: 66 of 75 (88 %). The difference is not statistically significant ($p = 0.15$), and the effect size is small ($d = -0.25$).

In the PyTamaro group, several participants changed numbers incorrectly in at least one place. In three cases, `triangle`'s angle argument was also doubled; in one case only the floor was updated, and in one other case only the roof; one participant multiplied numbers by 1.5 instead of 2; three other participants otherwise altered the numbers incorrectly; and one tried to inject an extra instruction in the loop body. Four participants did not submit a solution.

In the Turtle group, we expected to see several failures to update `forward`'s argument (as described in Section 3.6.5 above). However, of the nine incorrect solutions, only two were in this category. The other wrong solutions were either not submitted (four cases), had incorrect numerical arguments (two), or inadvertently redefined the name `square` (one).

### 4.4 Engagement and Perceived Learning (H3)

Table 8 summarizes the post-survey results, first comparing the groups and then looking at possible gender differences. Each claim, in order, tests the corresponding hypothesis (H3a to H3g).

■ **Table 8** Post-survey results divided by experimental group (left) and gender (right)

| Claim (Likert scale from 1 to 7) | PyT. | Tur. | p | d | Fem. | Male | p | d |
|---|---|---|---|---|---|---|---|---|
| *Engagement:* | | | | | | | | |
| I found the preceding lessons interesting | 6.06 | 6.00 | 0.74 | 0.06 | 6.08 | 5.95 | 0.46 | 0.13 |
| Programming with graphics is fun | 6.01 | 5.83 | 0.33 | 0.17 | 5.97 | 5.86 | 0.56 | 0.10 |
| I like programming with graphics more than the text-based programming we have done in the course | 4.44 | 4.56 | 0.66 | -0.08 | 4.73 | 4.29 | 0.11 | 0.30 |
| I would like to learn more about programming with graphics | 6.09 | 5.87 | 0.16 | 0.24 | 5.97 | 6.00 | 0.85 | -0.03 |
| *Perceived learning:* | | | | | | | | |
| I feel that I learned about programming concepts from these lessons | 6.07 | 5.64 | 0.03 | 0.36 | 5.82 | 5.85 | 0.88 | -0.03 |
| I already knew beforehand how to do graphical programming similar to what was taught | 1.64 | 2.29 | 0.02 | -0.39 | 2.03 | 1.88 | 0.60 | 0.09 |
| I already knew beforehand all the general programming content | 5.26 | 5.33 | 0.83 | -0.04 | 5.29 | 5.28 | 0.96 | -0.01 |

Overall, the participants liked the activities and found them useful for learning. PyTamaro users were more inclined to say that they "had learned about programming concepts from the lessons" ($d = 0.36$, $p < 0.03$). As expected, PyTamaro's compositional graphics approach was less familiar to the average participant than Turtle ($d = -0.39$, $p < 0.02$). Moreover, we observed small effects in favor of PyTamaro with respect to how much participants now enjoyed programming with graphics ($d = 0.17$) and their desire to continue with graphics-based programming ($d = 0.24$).

Differences between female and male students were minimal, with one exception. The participants' CS1 course relies on traditional programs with text-based console I/O. After the intervention, female participants were more likely to say that they like programming with graphics more than what they have done in CS1 (small-to-medium effect, $d = 0.30$).

Participants across groups largely agreed that "programming with graphics is fun" (average 5.9). This item is not identical to our more generic "programming is fun" item on the pre-test (average 5.4). With that caveat in mind, we note that there is a statistically significant difference between these within-subjects ratings, with a moderate effect size ($d = 0.46$, $p < 0.001$).

## 5  Discussion

In summary, we found that (1) both graphics-based approaches—compositional and turtle—engaged students and led to high ratings of perceived learning; (2) both groups performed uniformly well on the programming tasks in the post-test, except for a substantial difference in code tracing; and (3) the groups performed equally on post-test questions on conceptual knowledge. Sections 5.1 and 5.2 below elaborate on the first two points, respectively. In Section 5.3, we discuss the third point in detail as we consider both the factors that may have affected our specific result and, more broadly, the methodological issues that complicate studies such as ours.

### 5.1  Student Engagement Was High

Both the PyTamaro group and the Turtle group reported high levels of engagement with programming using graphics. They expressed enthusiasm for it, thought it was more likable than traditional non-graphics-based programming, and felt that they had learned programming concepts by engaging with it (Section 4.4). Our study thus adds to the body of evidence (Section 2) on the value of graphics in introductory programming education. Moreover, PyTamaro's compositional graphics approach appears to yield engagement levels at least on par with the venerable and widely popular turtle approach.

Learner diversity is one of the motivations to introduce graphics in CS1, and pedagogies such as media computation have had positive effects on gender balance [22, 23]. Engaging and retaining diverse students is also an explicit goal for PyTamaro, and two of our results are potentially significant in this light. First, almost half (49 %) of the volunteering participants identified as female, which is substantially higher than the proportion of female students in the CS1 course (ca. 35 %). We had advertised the study as "learning to program graphics in Python," which may have piqued female students' interest in particular. (It is known that there are differences in how female and male students value different domains in programming tasks [35].) Second, female participants especially agreed with the post-survey claim that they prefer programming with graphics to text-based programming (average 4.7 vs. male students' 4.3). This effect was of a small-to-moderate size but did not reach statistical significance ($d = 0.30$, $p = 0.11$). Our experiment was not designed to provide data on long-term impact, but our results do suggest that choosing graphics as a domain for introductory programming may have an immediate impact on stimulating female students' interest.

## 5.2 Differences Between Groups Were Scarce, With One Exception

Overall, the two experimental groups showed comparable results on the post-test tasks. We start the discussion with the notable exception of the performance difference on the tracing task, which tested hypothesis H2a.

### 5.2.1 The PyTamaro Group Did Better on Their Tracing Task

One of the goals of compositional graphics approaches is to encourage decomposing a problem into independent subproblems so that one may reason about each subproblem in isolation. This is harder to achieve with turtle graphics, as the turtle carries a state that includes its position and heading [10]. Tracing a turtle program requires one to track the state at each step through a sequence of commands.

While designing the tracing question (Section 3.6.3), we considered several aspects in an attempt to ensure a fair comparison. First, both groups used a `square` function they had already practiced during the intervention. Second, the `square` function we gave the Turtle group 'behaves nicely': it leaves the turtle facing in the same direction. Third, the PyTamaro code is suboptimal in several ways, compared to the code quality one would typically have with PyTamaro. (Specifically: Each variable is used only once. The `square` function is called four times merely to match the four calls in the Turtle program. Similarly, repetition of the first three lines could have been avoided by reusing `c`. The side length is not given a meaningful name; the literal `10` is passed directly as an argument. All the variable names are devoid of meaning.)

Our participants performed vastly better on the PyTamaro tracing task than on the Turtle task. This result comes with caveats. For one thing, there was only one tracing question for each participant. For another, the two groups traced different programs. Although we have argued that the two programs are, in a sense, comparable, our reasoning may be called into question. Nevertheless, the very large effect size in favor of PyTamaro (Section 4.3.1) suggests that tracing typical turtle graphics code requires more effort and care compared to typical compositional graphics code, even for non-novices. It is important to bear in mind that this finding speaks partially—and perhaps mainly—of the nature of the two programs and their associated learning approaches rather than between-group differences in learning gains; however, that finding, too, is relevant to instructors that employ graphics-based pedagogies.

### 5.2.2 Other Differences Were Largely Absent

We found no major differences between the groups on the code-writing task or the code-modifying task (Sections 4.3.2 and 4.3.3). By and large, both groups performed rather well on these tasks. Despite this result, the possibility certainly remains that the two approaches to programming graphics lead to differences in students' programming skills. However, it may be that observing such differences would require a context where learners practice with a graphics library longer (e.g., over several weeks) and can then demonstrate their abilities on significantly larger exercises. For us, arranging for such an experiment was not feasible in practice; it would have also reduced the degree of control over the study participants, likely introducing new confounding factors.

Similarly, we observed no substantial differences between the groups' performance on the six post-test questions about expression-related programming concepts (Section 4.2). Again, the short length of the intervention may have contributed, but there are many other factors worth considering as well, as we discuss below.

## 5.3 Question Design and Conceptual Transfer

Certain programming concepts are prominent in the compositional graphics approach that PyTamaro supports; these include nested function calls and other composite expressions, functions that take various numbers of parameters, and return values, to list a few. In the absence of a validated assessment instrument that focuses on these concepts, we designed an ad hoc one: questions Q1 to Q6 in our post-test (Appendix D). We applied this instrument in hopes of showing conceptual transfer from PyTamaro to a non-graphical domain (using turtle graphics as a baseline) but did not find evidence of it.

When designing and applying such an instrument, there are many decisions to be made. Below, we discuss a few of them in order to explain our design, reflect on possible reasons for the lack of observed differences between groups, and, perhaps, to highlight some pitfalls to others engaged in evaluative computing education research.

### 5.3.1 Staying Clear From 'Teaching to the Test'

We wanted at least part of the post-test to be identical for both groups. One reason for this was to enable direct comparisons between the groups on the same questions.

Identical questions do not guarantee a fair comparison, however. An inherent problem in evaluating educational innovations is that researchers may unwittingly favor a particular group—especially if they designed the innovation being studied. At an extreme, one group might be taught precisely what the post-test asks. Crichton and Krishnamurthi recently reflected on this while designing interventions to improve a textbook alongside assessments to evaluate the interventions: "If an intervention is too tailored to the specific question being targeted, then learners are likely not forming a robust mental model. We managed teaching-to-the-test by ensuring that interventions did not change the textbook to trivialize the problems under question, e.g., by adding the answer verbatim to the book." [13]

Not providing the exact answers to one group only is a start, but insufficient. If one group's intervention materials closely match the post-test, any success on the test might be mere memorization. In the words of Perkins and Salomon, "any learning requires a modicum of transfer. To say that learning has occurred means that the person can display that learning later." [47] In an attempt to capture genuine learning, we designed conceptual questions that were 'at a distance' from what the participants experienced during the intervention. We achieved this using a different domain than the graphic one which was used in the teaching materials, staying clear from the risk of 'teaching to the test.'

### 5.3.2 Transfer to Isomorphic Programs

Within the constraints of a short experiment, one can hardly expect *far transfer* to very dissimilar problems; there is plentiful evidence that far transfer is a lofty goal in general (e.g., [16, 28]). We instead aimed to find *near transfer*, asking the participants to answer questions about programs that are not in the domain of graphics but that are *isomorphic* with what the PyTamaro group was previously taught in that domain.

The diagram in Figure 2 illustrates how transfer to an isomorph doesn't happen 'directly.' Instead, learners are supposed first to acquire an 'abstract' understanding and then to apply it in a new context.

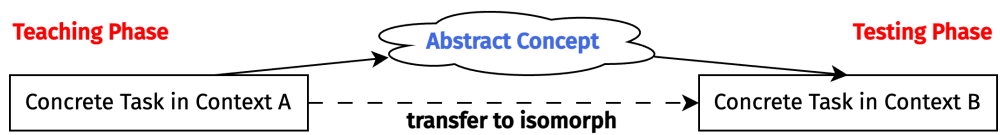

**■ Figure 2** Learners are exposed to a concrete task during the teaching phase. During the testing phase, they are assessed on an isomorphic task in a different context. Achieving transfer requires acquiring an understanding of the abstract concept.

We designed the multiple-choice questions with this idea in mind, focusing on the 'abstract concepts' that PyTamaro was hypothesized to teach (cf. Table 3, that shows the isomorphisms between the PyTamaro intervention and the post-test).

### 5.3.3 Transfer, Even to Isomorphic Tasks, Can Fail

Contrary to our hypothesis H1, the PyTamaro group, despite having practiced on tasks isomorphic to those in the post-test, did not perform better than the Turtle group. It is well known that transfer is not easy to achieve or to demonstrate through research. The result of this study provides further evidence that programming is no exception in this regard, even when aiming only for near transfer to isomorphic tasks.

One reason why even near transfer often fails is that learners—novices especially—may fixate on the surface characteristics of a task and consequently fail to draw the appropriate connections between tasks and abstract from them. This challenge has been noted, among others, by Perkins and Salomon, who wrote: "Subjects usually do not recognize the connection between one isomorph and the other and hence do not carry over strategies they have acquired while working with one to the other" [47]. Given that our post-test questions were in a different domain, albeit isomorphically so, it is debatable whether we truly tested for near or far transfer.

Transfer is likelier if instruction highlights the relationships between concrete tasks and abstract concepts. For instance, Reed et al. [50] discovered that many people would not transfer from the 'Jealous Husbands' problem to the similar 'Missionary–Cannibal' without explicit instruction. Our teaching interventions did not consistently and explicitly highlight opportunities for transfer, which may have affected our results. An improved intervention could embrace more deeply the idea of "semantic waves" [14, 36] and guide learners from the abstract to the concrete and then back to the abstract, highlighting the relationships between the two levels of abstraction.

## 6   Threats to Validity

**Prior Knowledge**   As discussed in Section 3.1, the lack of a pre-test is justifiable given the randomization and the large sample, and even has some advantages. Nevertheless, this decision does preclude us from computing 'learning gains,' and there remains the issue of how prior knowledge affected our participants' performance.

We targeted students who did not have much experience, but not all participants were novices. However, even with only novices included in the analysis, the differences between the two groups were minor. One plausible and perhaps likely explanation for this is that the participants' performance was sufficiently influenced by what they had previously learned in CS1 that any effect of either group's short intervention on the post-test is negligible in comparison.

The participants' CS1 course adopts a 'typical imperative' view of programming (loops, etc.) that is more closely attuned to the turtle approach than to PyTamaro's compositional graphics approach, as the latter instead emphasizes the compositional power of expressions that comes with 'functional' programming (nested calls, etc.). We speculate that this might have introduced a slight bias in favor of the Turtle group. Similarly, the Turtle group could have been advantaged by prior experience in 'computational thinking' or 'problem solving' activities that do not involve programming as such but that match the spirit of turtle graphics (e.g., [7]); such activities are not uncommon in secondary education.

**Study Duration**   The experiment was designed to allow participants to work through a programming tutorial with graphics for one hour. As noted in the discussion, the short duration enabled controlling certain variables (e.g., avoiding participants discussing the lessons' content between groups, having the exact teaching materials available for reproducibility), but also significantly limited opportunity to observe effects that are more visible in the medium and long term.

**Data Collection**   Engagement and prior experience were self-reported and potentially subject to self-reporter bias.

The items for our surveys and post-test were constructed ad hoc and not validated. Several post-test items were in the multiple-choice format, which can be problematic if the options are not paired with explanations [8]. We mitigated this by including an "I don't know" option, but our questions may nevertheless have been too 'guessable.'

The items on tracing, modifying, and writing code were designed to be 'comparable' between groups but not identical.

**Generalizability**   All our participants come from a single university course, albeit one with students from a wide variety of engineering majors; we cannot comment on how our results might generalize to other groups. Moreover, our subjects were volunteers who received a small compensation for participating, which may have introduced a selection bias.

**Response Bias**   We solicited answers from participants both in the pre- and the post-survey. Participants did not engage with a 'teacher,' but may still have skewed their answers due to the social-desirability bias. In particular, the pre-survey asked participants to voluntarily disclose their gender, which is known to possibly affect their behavior (for example with respect to the stereotype threat).

**Authorship Bias**   We preemptively compensated for an authorship bias when designing the experiment, disadvantaging our own library in several ways. (1) We placed PyTamaro in an imperative-style CS1 that does not play to PyTamaro's strengths. (2) We used animated (thus possibly slightly more engaging) visuals only in the Turtle materials, as we thought they were needed to provide a high-quality explanation of the turtle's state changes over time. (3) We introduced to both groups the idea of a loop to accumulate a value, despite it not normally being covered in Turtle pedagogy, as we felt it necessary to fairly prepare all participants for one question on the post-test. (4) We wrote the tracing question in a style far from optimal for PyTamaro programs (Sections 3.6.3 and 5.2.1). (5) We took care of setting up the turtle's drawing environment (e.g., providing a large enough canvas).

   Ultimately, we cannot rule out an authorship bias. For transparency, the complete materials used in the study are available as appendices.

## 7   Conclusion

In this article, we have presented a randomized, controlled experiment on the use of graphics in teaching programming to beginners. We found that both a compositional graphics approach enacted using the PyTamaro library and a more traditional turtle graphics approach engaged student programmers; female students might find such approaches particularly engaging. We did not find evidence of better transfer from a short PyTamaro session to a post-test on isomorphic tasks outside the graphics domain, compared to the Turtle session which did not feature those tasks. Overall, there were few differences between the two experimental groups. As an exception to that trend, beginners appear to trace compositional graphics code more accurately than 'comparable' turtle graphics code.

   We have outlined several alternative—or complementary—explanations for our findings. Further research is needed to test our speculations as well as the generalizability of our results. Ultimately, our findings are inconclusive regarding compositional graphics approaches such as PyTamaro's; we recommend that future research look into interventions longer in duration than what we were able to investigate here.

**Acknowledgements**   We thank the reviewers for their observations that helped clarify several parts of this article. We are grateful to the instructors of the Y1 course at Aalto for allowing us to advertise this study, and to the students who participated.

   This work was partially funded by the Swiss National Science Foundation project 200021_184689.

### A   Pre-Survey

Participants answered a pre-survey with the following questions:

- On demographics:
  - *How old are you?* [numeric]
  - *What is your gender?* [male; female; non-binary; other; prefer not to answer]
- On prior experience:
  - *How many lines of code have you written before starting the course, across all languages except HTML and CSS?* [none; fewer than 50; fewer than 500; fewer than 5000; more]
  - *Which rounds of exercises of the course have you completed?* [subset of #1, #2, #3]
  - *Have you ever written a program that draws graphics before?* [yes; no; not sure]
- On the attitude towards programming (answers on a seven-point Likert scale):
  - *It is useful for me to know how to program.* [seven-point Likert from "not at all true" to "completely true"]
  - *Programming is boring.* [seven-point Likert]
  - *Programming is fun.* [seven-point Likert]

### B   Teaching Intervention

This Section contains the exact text of the teaching intervention used in the study. The intervention was divided into four mini-lessons, which correspond to the four subsections below. Each lesson starts with a *common part* that was shown to both the PyTamaro and Turtle groups. The lesson then "splits in two": participants in the PyTamaro group worked through the part marked as "PyTamaro-Only", whereas participants in the Turtle group worked through the part marked as "Turtle-Only". A short recap concludes each lesson and is common to both groups.

During the intervention, the Python code were shown in a web-based environment that allowed participants to run the code and see the output.

#### B.1  Mini-Lesson 1 (of 4)

All the programs you have written so far in the CS1 course deal with text: perhaps they read some input from the user as a sequence of characters (that is, a string), they do some processing and calculations, and call the function `print` to spit out an answer that is again textual.

Over the course of the next hour, you will learn how to write programs that go beyond that and can create graphics. If that sounds scary, fear not!

**Libraries**   It is much quicker to reuse program code that someone has already written, instead of starting from scratch. This is why programmers constantly use so-called **libraries** of reusable program parts to accomplish various things.

You can think of a library simply as a collection of **names**. Some of those names refer to **functions** that the library provides for you to use: for instance, Python's `math` library offers a function named `sqrt` which computes the square root of a number. Other names might just refer to plain **values**: again as an example, Python's `math` library contains the name `pi` for the mathematical constant $\pi$, and thus `pi` equals approximately 3.1415.

The tiny Python program below prints an approximate value of `pi`, which comes from the `math` library. Note that you can choose to run the program — try it!

```
1  from math import pi
2  print(pi)
```

The first line of the program above "imports" the name `pi` so that you can use it in your program. While the import is a necessary step, you will not see it again in all the programs featured in the rest of these less material. You do *not* need to worry about that: we add all the necessary imports automatically for you behind the scenes.

**There are many libraries; we'll use one of them**   When they need to perform a task, programmers can decide to use one of the many available libraries. Drawing graphics is no exception: there are many available libraries to draw onscreen, and they embrace different approaches.

This study explores two different libraries to draw graphics in Python. You have been randomly assigned to one of them, which you are going to use in this session. After the study is over, however, you are also welcome to look at the materials for the other library, if you are interested.

<div style="background:blue;color:white;text-align:center;font-weight:bold;">PyTamaro-Only Part</div>

**Let's Draw a Rectangle**   Let's dive into it and see how to use a library named **PyTamaro** to draw graphics. We start very humbly, writing a program to create a rectangle and show it on the screen. Conveniently, the PyTamaro library offers you a function named `rectangle` to create rectangles. The library also has names such as `green` for basic colors.

How do you call the `rectangle` function? Here is an illustration of the basic idea:

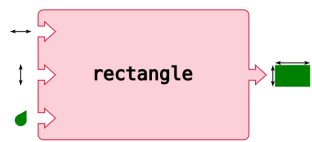

■ **Figure 3**   Example call of `rectangle`

`rectangle` takes in three **parameters**, which are represented in the above illustration by the three incoming arrow-shaped "holes". The first two parameters determine the width and the height of the rectangle; the third one determines the color.

We can use numbers such as `200` and `100` for the first and the second parameter, to indicate the width and the height, and a color such as `green` for the third parameter.

How can we use this function in a Python program? Let's store its **return value**, our rectangle (represented in the image near the outgoing arrow), in a variable named `football_field`. (There's an example of this below.)

And then, just like one can pass any string such as `"Hello world"` to the function `print` to display a text on the screen, let's pass a graphic to a function named `show_graphic` to display it onscreen.

Try it! In the code below, the `show_graphic` function call contains three dots (`...`). Edit the program: replace the dots with the name of the variable where we stored the rectangle. Then run the program.

```
1 football_field = rectangle(200, 100, green)
2 show_graphic(...)
```

Have you managed to see your first graphic? Congratulations!

**Let's Rotate Things**  Imagine now that you are sitting right at a corner of a stadium: the football field would not look to you "horizontal", but rotated by some angle. We can try to modify our program to draw something like that. The PyTamaro library offers you a function named `rotate` that takes two parameters: an angle in degrees and a graphic. The function returns a graphic rotated counterclockwise by that angle. For example:

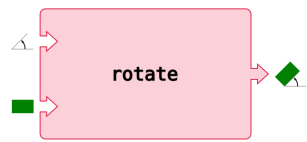

■ **Figure 4**  Example call of `rotate`

In the next program below, complete the assignment in the second line by replacing the dots with a call to the function `rotate`. For `rotate`'s first parameter, write `45` (i.e., 45 degrees); for the second, write `football_field`. Do not forget to separate the two parameters with a comma.

```
1 football_field = rectangle(200, 100, green)
2 rotated_field = ...
3 show_graphic(rotated_field)
```

Do you see a rotated field? Awesome!

Perhaps you are wondering if we really *must* have two variables in the program above. The answer is: no. We can rewrite the solution to combine the function calls to `rectangle` and `rotate`. In terms of the illustrations above, we are plugging the outgoing arrow of `rectangle` straight into the second incoming "hole" of `rotate`.

Take another look at the program just above. We need to plug the call to `rectangle` into the place reserved for the second parameter when we are calling the function `rotate`. Do that now in the code below: copy the expression

`rectangle(200, 100, green)` (the entire expression, including the closing `)`) and paste it below, replacing the three dots `....`

```
1 rotated_field = rotate(45, ...)
2 show_graphic(rotated_field)
```

<div align="center">

**Turtle-Only Part**

</div>

**Let's Draw a Rectangle** Let's dive into it and see how to use a library named **turtle** to draw graphics. We start very humbly, writing a program to create a line to be shown on screen.

The turtle library is based on the following metaphor.

Imagine you are controlling a "robotic turtle" that can move on a canvas carrying a colored pen. You give commands to the turtle. When the turtle moves, it leaves a trace on the canvas, ultimately producing a drawing.

One of the commands understood by the turtle is `forward`. The function `forward` takes in one number as a **parameter**. Calling the function causes the turtle to move forward by the given amount of steps.

Note that `forward` moves the turtle in the direction it is *currently facing*. The turtle starts facing east (that is, towards the right of your screen). Here is an animation of the turtle moving `forward` with `100` steps:

■ **Figure 5** Animation forward [last frame — only the last frame of animations is reproduced in this article, but the full animation was visibile to the participants during the study]

We can also change the color of the pen carried by the turtle, to draw colored lines such as in the example above. The default pen color is black, but we can use the function `pencolor` to use a differently colored pen.

The `pencolor` function takes in one single parameter, a string containing the name of the desired color for the pen.

To recap, you can draw a colored line by calling `pencolor` with a string parameter such as `"green"` for the name of the color, and further calling `forward` with a numerical parameter such as `100` for how much the turtle should move forward.

Try it by yourself! Replace the three dots `...` inside the call to `pencolor` with the appropriate string to draw a green line.

```
1 pencolor(...)
2 forward(100)
```

Have you managed to see your first graphic? Congratulations!

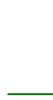

■ **Figure 6**  Animation forward-left [last frame]

**Let's Rotate**  We cannot get very far just by commanding the turtle to move forward. Luckily, there are functions that rotate the turtle — or, in other words, change the direction that the turtle faces in. Two such functions are named `left` and `right`. For example, the `left` function offered by the library has one single parameter: an angle in degrees that indicates how much the turtle should turn *left*.

We can draw a "mirrored" letter L (something like ⌐) by first moving forward as before (with the turtle moving towards the right edge of the screen), then turning the turtle left by 90 degrees, and finally moving forward again.

Here is an animation that shows the plan for our turtle:

In the third line of the following program, replace the dots with a call to the function `left` as described above.

```
1 pencolor("green")
2 forward(100)
3 ...
4 forward(200)
```

Do you see something that resembles a ⌐? Awesome!

Perhaps you are wondering if we cannot draw a proper letter L. The answer is: we can. To do so, we'll make use of the `backward` function. The function behaves exactly like `forward` but moves the turtle backward (relative to the direction it is facing).

Here is the plan to draw the letter L. As always, the turtle starts facing east. We move forward by a certain amount, and then we move backward by the same amount. Then, as before, we turn left 90 degrees and move forward to complete the letter.

Now, in the code below, replace the three dots with `backward(100)` to complete the plan above.

```
1 pencolor("green")
2 forward(100)
3 ...
4 left(90)
5 forward(200)
```

> **Common Part (PyTamaro and Turtle)**

**So Far, So Good!**  If everything worked, give yourself a pat on the back! In this lesson, you learned what a programming library is and how to use one to draw a very simple graphic. On to the next adventure!

## B.2 Mini-Lesson 2 (of 4)

You now know how to draw an extremely basic shape. In this lesson, we will step up the game a bit and try to draw a house. The house is made up of a ground floor, represented by a square, on top of which sits a roof, represented by an equilateral triangle.

<div style="background:#3b5bdb;color:white;text-align:center;font-weight:bold;padding:6px">PyTamaro-Only Part</div>

**Let's Draw the Ground Floor**  You can use the `square!` function to create the ground floor. It takes two parameters, the side length and the color, and returns a graphic of a square.

In the program below, replace the dots on the first line. The line should assign value to the variable `ground_floor`: the should variable holds the return value of the `square` function, when that function is called with `100` as the side length and `yellow` as the color.

```
1 ground_floor = ...
2 show_graphic(ground_floor)
```

(The `square` function works by creating a rectangle of the specified color in which width and height are the same. In fact, it internally makes use of the `rectangle` function seen in the previous lesson.)

**Let's Draw the Roof**  We now need to figure out how to create the roof. Luckily, the PyTamaro library offers a function named `triangle`. It takes four parameters. The first two determine the lengths of two of the triangle's sides; the third parameter determines the angle between those sides. The last parameter, as usual, describes the color.

Let's create a red equilateral triangle with a side length of 100. All of an equilateral triangle's internal angles measure 60 degrees. Therefore, we can call the `triangle` function passing in the values `100`, `100`, `60`, and `red`. Visually:

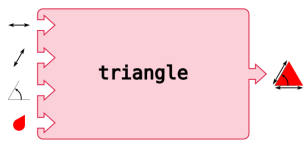

■ **Figure 7**  Example call of `triangle`

Replace the dots in the code below with the appropriate function call.

```
1 roof = ...
2 show_graphic(roof)
```

Great! Now that we have the two individual graphics, we need a way to combine them together as we intend.

**Let's Put the Pieces Together**   PyTamaro caters to our needs: it offers a function `above` to place two graphics one above the other. The function `above` places the graphic received as the first parameter above the graphic specified by the second parameter. It returns a new, composed graphic.

Here is how we intend to use it:

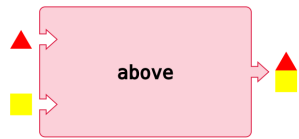

■ **Figure 8**   Example call of the `above` function

Complete the code below by replacing the dots in the third line with a call to the function `above`, passing in the two graphics that are stored in the `ground_floor` and the `roof` variables.

```
ground_floor = square(100, yellow)
roof = triangle(100, 100, 60, red)
house = ...
show_graphic(house)
```

Once you see the house, you can also experiment and exchange the two parameters in the call to `above`. Run the program again and observe the difference!

<div align="center">**Turtle-Only Part**</div>

**Let's Draw the Ground Floor**   You can use the function `square` to draw the ground floor. It takes just one parameter, the side length.

It is useful to understand just how this `square` function works. It repeats this combination of commands four times: 1. it moves the turtle forward by the given side length, and 2. then it rotates the turtle 90 degrees to the right.

Remember that *at the beginning* of a program the turtle starts by facing east (that is, towards the right of the screen). Assuming that as a starting point, this is what happens when the function `square` is called: 1. The turtle moves forward drawing the top side of the square, and then it turns right. At this point, the turtle is facing south. 2. The turtle moves forward drawing the right side of the square, and then it turns right. At this point, the turtle is facing west (that is, towards the left of the screen). 3. The turtle moves forward drawing the bottom side of the square, and then it turns right. At this point, the turtle is facing north. 4. The turtle moves forward drawing the left side of the square, and then it turns right.

After all the steps, the turtle has drawn a square. It is again facing in the same direction as before the execution of the `square` function. The animation below exemplifies this process.

Now that you have an idea of how the `square` function works, replace the dots with a call to it, using `100` as a value for the first and only parameter.

```
pencolor("yellow")
...
```

■ **Figure 9**   Animation, drawing a square with turtle [last frame]

**Let's Add the Roof**   We now need to figure out how to create the roof. For that, we can use a function named `triangle`, which is similar to `square`.

   `triangle` commands the turtle to move forward and turn right 120 degrees *three times*. In practice, this means that the function draws the three sides of a triangle one after the other. The animation below shows what happens when the `triangle` function is called with a side length of `100`. This animation, too, assumes that the turtle faces east just before calling the function.

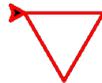

■ **Figure 10**   Animation, drawing a triangle with turtle [last frame]

   In the code below, replace the dots with a call to `triangle`, passing in the value `100` for the side length. Observe that before that command we have already added a call to `pencolor`, so that the roof is drawn in red.

```
1 pencolor("yellow")
2 square(100)
3 pencolor("red")
4 ...
```

   Whoops! You might have seen a drawing that was not what we intended! Don't worry: there is a simple explanation. When we call `triangle`, the first thing that happens is that the turtle moves forward, in order to draw the first side of the triangle. But this happens after the call to `square`, with the turtle (again) facing east; calling `triangle` makes it move forward — east — and then turn 120 degrees to the right, which is not what we want.

   We need to rotate the turtle before calling the `triangle` function, so that the turtle well positioned to start drawing the roof simply by moving forward.

   We know how to do that: we can use the `left` function to turn the turtle left by 60 degrees (the measure of an equilateral triangle's internal angle) just before drawing the triangle.

   Replace the dots in the code below with the appropriate call to `left`.

```
1 pencolor("yellow")
2 square(100)
3 pencolor("red")
4 ...
5 triangle(100)
```

| Common Part (PyTamaro and Turtle) |
|:---:|

Wow! That took a while, but you can now feel proud: you wrote a program to draw a house!

### B.3 Mini-Lesson 3 (of 4)

Our single house feels lonely: it is time to give it company.

**Two Houses**    Let's draw a duplex house, which simply means *two* single houses next to each other.

| PyTamaro-Only Part |
|:---:|

The PyTamaro library offers a convenient `beside` function that works pretty much like the `above` function you have already used.

It takes two graphics as parameters and returns a new graphic by placing one specified as the first parameter on the left, and the one specified as the second parameter on the right.

Complete the code below by replacing the dots with a call to `beside`. As you call `beside`, you can use the variable `house` *twice*, once for each parameter (since we're placing two identical houses side by side).

```
1 ground_floor = square(100, yellow)
2 roof = triangle(100, 100, 60, red)
3 house = above(roof, ground_floor)
4
5 two_houses = ...
6 show_graphic(two_houses)
```

Got two houses? Great! However, some privacy is always welcome.

**Privacy, Please!**    Could we add a wall in between the houses? A narrow, black rectangle will do the job.

But how? We'd like to place *three* graphics next to each other: a house, a wall, and another house. But what we have is a *two*-parameter function `beside` that places *two* graphics next to each other.

Well, we can call our function twice: the first call can combine the left house with the wall, while the second one can combine the previous result and the right house.

Implement this idea in the code below, replacing the dots with the appropriate calls to `beside`:

```
1 ground_floor = square(100, yellow)
2 roof = triangle(100, 100, 60, red)
3 house = above(roof, ground_floor)
4
5 wall = rectangle(15, 187, black)
6
7 left_house_with_wall = ...
8 two_houses_with_wall = ...
```

```
 9
10  show_graphic(two_houses_with_wall)
```

**An Alternative Solution**   There is a different but equally valid solution that perhaps already occurred to you.

There's no rule saying that the first step *must* involve the left house and the wall. We could first combine the wall and the right house so that they are next to each other, and then combine the left wall with the previous result!

Try to implement this variant and verify that it actually produces the same drawing.

```
 1  ground_floor = square(100, yellow)
 2  roof = triangle(100, 100, 60, red)
 3  house = above(roof, ground_floor)
 4
 5  wall = rectangle(15, 187, black)
 6
 7  wall_and_right_house = ...
 8  two_houses_with_wall = ...
 9
10  show_graphic(two_houses_with_wall)
```

<div align="center">

**Turtle-Only Part**

</div>

We can accomplish this with a rather simple idea. We have to: 1. Draw the first house. The turtle ends up at the top-left corner of the square. It is *not* facing east, however, because we rotated it 60 degrees left before calling `triangle`. 2. Re-position the turtle so that it will be ready to execute again the same commands as the first step. For this we need to: (a) compensate for the left turn made before calling `triangle` by turning right and (b) then move the turtle forward. 3. Execute the same commands as the first step to draw the second house.

Complete the code below with two appropriate lines that replace the dots with: 1. a call to `right` with `60` degrees as a parameter, to undo the left turn; and 2. a call to `forward` to move the turtle by the same width as one house, that is `100` steps.

```
 1  pencolor("yellow")
 2  square(100)
 3  pencolor("red")
 4  left(60)
 5  triangle(100)
 6  ...
 7  ...
 8  pencolor("yellow")
 9  square(100)
10  pencolor("red")
11  left(60)
12  triangle(100)
```

Note something important, though! When you moved the turtle forward during the second step, it is still carrying a red pen that draws. This turns out not to be a problem in this specific program, since the turtle moves along a line that has already been drawn in red. Going over it a second time does not do any harm. But be mindful of this pitfall in general: otherwise, your drawing may have surprising and unwanted lines.

So you got two houses? Great! However, some privacy is always welcome.

**Privacy, Please!**   Could we add a thick wall between the houses? A black square will do the job.

Modify the program below to add a wall, which requires replacing the dots with these three steps:

1. Change the color of the pen to `"black"`.
2. Draw a square of side 100.
3. Position the turtle appropriately so that it is ready to draw the second house.

Note that the third step is essential and that, in general, the order in which you give commands to the turtle matters.

```
1  pencolor("yellow")
2  square(100)
3  pencolor("red")
4  left(60)
5  triangle(100)
6  right(60)
7  forward(100)
8  ...
9  ...
10 ...
11 pencolor("yellow")
12 square(100)
13 pencolor("red")
14 left(60)
15 triangle(100)
```

(Do not worry about the left border of the second house overwriting the black wall.)

Once you got the proper drawing, convince yourself that the order of the commands matters. Suppose that the plan above had step 1 and step 2 swapped (which means drawing the wall before changing the pen color). Modify the code above to reflect this change. Observe the result: what happens to the drawing?

<div style="background:#4864d8; color:white; text-align:center; font-weight:bold; padding:6px;">Common Part (PyTamaro and Turtle)</div>

You have practiced drawing slightly bigger graphics. Let's bring this one step forward with the next lesson.

## B.4  Mini-Lesson 4 (of 4)

Let's build some bigger graphics!

In many pictures, there are *repeated* elements; a picture might have several houses, for example. That does not mean we have to duplicate a lot of code! The computer is excellent at repeating things for us.

In the CS1 course, you have already encountered one mechanism in Python to *repeat*: the `for` loop. Before getting back to graphics, let's review a simple example that might help you to refresh your knowledge.

**Recap: for Loops**    Suppose you want to write a simple program to compute the average of the five grades you obtained in the past semester. Each grade is asked to the user using the function `input`.

To compute the average, we need to divide the sum of all grades by the number of grades. How do we keep track of the sum of all grades? We can use a variable, named for example `sum_grades`. At any point during the execution of the program, the role of that variable is to track the sum of the grades seen so far.

```
1  n_grades = 5
2
3  sum_grades = 0
4  for i in range(n_grades):
5      grade = int(input())
6      sum_grades = sum_grades + grade
7
8  average = sum_grades / n_grades
9  print("Average grade:", average)
```

The `for` loop repeats the instructions "contained" in it (the two indented lines) `n_grades` times, which here means five times. In other words, the loop does five *iterations* over the instructions; we'll use this term below.

Before the first iteration, we do not have any information about the grades, yet. We can conveniently initialize the variable `sum_grades`! to 0.

Now consider the first iteration. For example, let's say that the user first enters the grade 4. Then, `sum_grades` will be assigned to the value `0 + 4`, which is just 4.

Imagine that the user inputs the grade 3 at the second iteration of the loop. The value of the variable `sum_grades` will be updated to `3 + 4`!, that is 7.

This process goes on for all the specified number of iterations (`n_grades`, in the example). At the end of the last iteration, the variable `sum_grades` has *accumulated* the sum of all grades, exactly like we wanted. We can then easily compute and print the average.

**Back to Graphics!**    Can we use a `for` loop to draw a graphic containing a repeated pattern? Sure we can!

Consider a simplified street that consists of a number of houses, all having the same appearance as the one we have drawn so far. We are looking at a densely populated neighborhood: there is no space between adjacent houses.

**PyTamaro-Only Part**

We can make good use of the `for` loop to *repeat* the same operation multiple times and place many graphics next to each other.

In the example presented at the beginning of this lesson, we used a variable (`sum_grades`) to *accumulate* the grades summed at each iteration. We can do the same and use a variable to accumulate the houses placed next to each other at each iteration.

Just like for grades we added *one* new grade at each iteration, we will now add *one* new house at each iteration to the ones "joined" so far.

Let's give the name `street` to the variable used to accumulate the houses. Before the loop, our street is going to be empty (of houses and indeed of anything). The first iteration will update `street` so that it contains one house. The second iteration will add one more house, so that `street` will contain two houses. The third iteration will add one more house: `street` will then contain three houses, and so on.

Which initial value should we use for `street`, before the loop? Like `0` in the example with the grades, we should use a value that works for the first iteration of the loop. Here, we need a graphic that when placed `beside` our first house, just results in that same single house.

PyTamaro has a function for this purpose: it is named `empty_graphic`. The function takes no parameters and returns an *empty graphic*. When combining an *empty graphic* with any other graphic (using `beside`, for example), the result is just the other graphic. Convenient, and a bit like zero in math!

Look closely at the code below. The first three lines create a house as we have always done so far. A variable `n_houses` contains the number of houses we want to have in our street. We initialize `street` to an empty graphic, the result of *calling* the parameterless function `empty_graphic`.

Then comes our `for` loop.

At each iteration, we need to assign to `street` a combined graphic. That graphic is the result of placing any previous houses beside one more house; in other words, we should place `street`'s earlier value beside a new house from the `house` variable.

Replace the dots below with a call to `beside`. As parameters, write the names of the two variables suggested above.

After the loop, `street` is a graphic that contains five houses next to each other, and is ready to be shown as usual with `show_graphic`.

```
ground_floor = square(100, yellow)
roof = triangle(100, 100, 60, red)
house = above(roof, ground_floor)

n_houses = 5

street = empty_graphic()
for i in range(n_houses):
    street = ...

show_graphic(street)
```

Can you see a street with five houses? Lovely!

**Turtle-Only Part**

We can now make good use of the *for* loop we just reviewed to *repeat* the same operation multiple times and place many graphics next to each other.

We need to draw a *street* of houses. Each iteration of the for loop will just draw *one* house.

Let's recall the plan we used in the previous lesson to draw just *two* houses: 1. Draw the first house. 2. Re-position the turtle so that it will be ready to execute again the same commands as the first step. 3. Draw the second house.

This implies that at the end of each iteration we need to prepare the ground so that the next one can start properly. Concretely, it means that after drawing a house (step 1 in the plan), we always need to: (a) turn the turtle right by 60 degrees (to "undo" the left turn made before drawing the triangle), and (b) move the turtle forward by 100 steps.

Complete the code below with the two appropriate commands so that after each iteration, the turtle is positioned so that it's ready to start drawing the ground floor of the next house.

```
1   n_houses = 5
2
3   for i in range(n_houses):
4       pencolor("yellow")
5       square(100)
6       pencolor("red")
7       left(60)
8       triangle(100)
9       ...
10      ...
```

Can you see a street with five houses? Lovely!

There is just one tiny inefficiency: we also reposition the turtle at the *last* iteration of the loop, even though there is no house to draw further. You can safely ignore this, given that we are not at all concerned with performance here.

As a final point, notice how using a loop helped us to avoid duplicating code, which is something that we did in the previous lesson, in which all the commands to draw a house were written twice. Experienced programmers consider code duplication a very bad thing. Think about what you would need to do if you had many houses in a graphic and decided that their roofs should be rectangles instead. It would require you to go through lots of lines and replace every occurrence of `triangle` with other code. Besides being a boring manual process, you would risk forgetting to do some replacements.

> **Common Part (PyTamaro and Turtle)**

**End of the Mini-Lessons**   You have now practiced for loops a bit more and learned how they help also in programs that deal with graphics.

## C   Post-Survey

Participants answered a post-survey with the following questions, all on a seven-point Likert scale from 1 ("not at all true") to 7 ("completely true"):

- I found the preceding lessons interesting.
- I feel that I learned about programming concepts from these lessons.
- I already knew beforehand how to do graphical programming similar to what was taught in the lessons.
- I already knew beforehand all the general programming content (variables, functions, loops, etc.) that was covered in the lessons.
- Programming with graphics is fun.
- I like programming with graphics more than the text-based programming we have done in CS1.
- I would like to learn more about programming with graphics.

### D  Post-Test Multiple-Choice Questions

For each multiple-choice question, participants have been asked to choose the claim they believe is most accurate. Questions featured an additional "I don't know" option, to be picked only in case the participant was very unsure.

#### D.1  Question 1

"Cha Cha Cha" is the title of a song. This Python program plays with the song title and prints "Cha" three times, each one on a separate line.

```python
print("Cha")
print("Cha")
print("Cha")
```

Your friend says that it is possible to get the same output differently by introducing a variable `word`:

```python
word = "Cha"
print(word)
print(word)
print(word)
```

Is the program still working as before?

- Yes, because `word` is used only once in each instruction/line. An instruction like `print(word + word + word)` is invalid and produces an error.
- Yes, because we can use the value stored in the variable as many times as we want.
- No, only the first `print` works. To fix the second program, we would need to add `word = "Cha"` before the second and the third `print` as well.
- No, because the second program prints three times `word`.

### D.2  Question 2

Python's math library contains a function named `sqrt`. It takes one parameter, the number to compute the square root of. Its return value is also a number, the square root of the provided number.

This program first computes the square root of 16, and then the square root of the result, which is finally printed:

```
root_of_sixteen = sqrt(16)
final_root = sqrt(root_of_sixteen)
print(final_root)
```

Your friend says that the same result can be obtained with a shorter program:

```
print(sqrt(sqrt(16)))
```

Is the program still working as before?

- Yes, because of the mathematical properties of the square root function. The same transformation with a function `half` that divides a number by two would not have worked.
- Yes, because it first computes the square root of 16, then computes the square root of the result, and eventually prints the final result.
- No. `sqrt(16)` works, because we are passing a number, 16, to the function. But in `sqrt(sqrt(16))` we are passing `sqrt(16)`, which is not a number but a function call.
- No, because `sqrt` is a function that takes one parameter, and the second program attempts to give the first (outermost) `sqrt` call two parameters.
- No, because we need a variable to store the result of `sqrt` before we can pass it to another function call.

### D.3  Question 3

Imagine that a Python library contains a function named `subtract`. It takes two numbers as parameters, and returns the result of subtracting the second number from the first one.

```
result = subtract(10, 7)
print(result)
```

Does executing the program above print 3?

- Yes, because calling `subtract` is one way to subtract a number from another. Because we are free to choose the order of parameter values, we could have also written `result = subtract(7, 10)` to get the same result.
- Yes, because we are correctly passing the numbers 10 and 7 to the function `subtract`.
- No, because functions can only have one parameter. It is therefore impossible for the library to offer a working `subtract` function with two parameters.
- No, because that is not how you should pass multiple parameters when a function requires more than one: the call should have been `subtract(10)(7)`.

### D.4 Question 4

Imagine that a Python library contains a function named `fake_random`. It has zero parameters, and always returns the number `42` as a fake random number.

```
1 print(fake_random())
```

Does executing the program above print `42`?

- Yes, because `fake_random()` calls the function, which will return the number `42`. The result is passed to `print`. Also, the empty parentheses `()` are necessary; just `print(fake_random)` does not work.
- Yes, because `fake_random()` calls the function. Also, writing just `print(fake_random)` without the empty parentheses would have done the same, given that the function returns a constant number.
- No, because the function `fake_random` cannot possibly exist as such, as functions need to have at least one parameter.
- No, because the function `fake_random` has zero parameters, and such a parameter-less function cannot return a value.
- No, because we need a variable to store the result of `fake_random` before we can pass it to `print`.

### D.5 Question 5

Imagine you have a function named `combine` at your disposal. It takes two strings as parameters and returns a combined string. For example, `combine("hel", "lo")` returns `"hello"`.

The goal is to write a program that constructs the word `restaurant` from three pieces, then prints out the result.

One of your friends comes up with the following program:

```
1 first_combination = combine("re", "stau")
2 word = combine(first_combination, "rant")
3 print(word)
```

Another friend suggests this other implementation:

```
1 first_combination = combine("stau", "rant")
2 word = combine("re", first_combination)
3 print(word)
```

What can you say about these two programs?

- They both work, even if the word is constructed in two alternative ways.
- The first one works, but the second one does not, because `"re"` is added in the second line after `"staurant"` has been created.
- The first one works, but the second one does not, because `combine` is not commutative (that is, because exchanging the first parameter with the second makes a difference).

- Neither one works, as a two-parameter function cannot be defined (so that it works).

### D.6 Question 6

You want to write a program that asks the user for five numbers, multiplies all of them together, and prints the result. Your friend suggests this skeleton, but they are unsure about what to write instead of the dots at the beginning.

```python
1 ...
2 for i in range(5):
3     number = int(input())
4     product = product * number
5 print(product)
```

What should the dots be replaced with?

- `product = 0`, because the variable `product` needs to be initialized to the neutral number `0` before looping.
- `product = 1`, because `1` is the only number that multiplied with any other number just results in the other number.
- We need to initialize `product` to some value, but it doesn't matter which one, because that value will in any case be replaced by the first number entered by the user during the first iteration of the loop.
- There is no number that works as an initial value for `product`. Other changes would need to be done to the program as well.

## About the authors

**Luca Chiodini** is a PhD student at the LuCE research lab at Università della Svizzera italiana. Contact: luca.chiodini@usi.ch.
https://orcid.org/0000-0002-2712-9248

**Juha Sorva** is a Senior University Lecturer in Computer Science at Aalto University. His research interests include programming education and instructional design. His other interests include Dr. Pepper Zero and the etymology of the word "aftermath." Contact: juha.sorva@aalto.fi.
https://orcid.org/0009-0003-1727-1317

**Arto Hellas** is a Senior University Lecturer at Aalto University. His current research interests include understanding and improving teaching and learning in digital and hybrid learning environments. He is also a fan of dad jokes. Contact: arto.hellas@aalto.fi.
https://orcid.org/0000-0001-6502-209X

**Otto Seppälä** is a University Lecturer at Aalto University. Contact: otto.seppala@aalto.fi.
https://orcid.org/0000-0003-4694-9580

**Matthias Hauswirth** is an Associate Professor leading the LuCE research lab at Università della Svizzera italiana. Contact: matthias.hauswirth@usi.ch.
https://orcid.org/0000-0001-5527-5931